\documentclass[pre,twocolumn,showpacs,amsmath,amssymb]{revtex4}
\usepackage[]{graphicx}

\begin{document}


\title{Dynamics of lane formation in driven binary complex plasmas}

\author{K. R. S\"utterlin$^1$, A. Wysocki$^2$, A. V. Ivlev$^1$, C. R\"ath$^1$, H. M. Thomas$^1$, M. Rubin-Zuzic$^1$,
W. J. Goedheer$^3$, V. E. Fortov$^4$, A. M. Lipaev$^4$, V. I. Molotkov$^4$, O. F. Petrov$^4$, G. E. Morfill$^1$, H.
L\"owen$^2$}

\affiliation{$^1$Max Planck Institute for Extraterrestrial Physics, 85741 Garching, Germany\\
$^2$Heinrich-Heine-Universit\"at D\"usseldorf, 40225 D\"usseldorf, Germany\\
$^3$FOM-Institute for Plasma Physics Rijnhuizen, 3430 BE Nieuwegein, The Netherlands\\
$^4$Joint Institute for High Temperatures, 125412 Moscow, Russia}

\begin{abstract}
The dynamical onset of lane formation is studied in experiments with binary complex plasmas under microgravity conditions.
Small microparticles are driven and penetrate into a cloud of big particles, revealing a strong tendency towards lane
formation. The observed time-resolved lane formation process is in good agreement with computer simulations of a binary
Yukawa model with Langevin dynamics. The laning is quantified in terms of the anisotropic scaling index, leading to a
universal order parameter for driven systems.
\end{abstract}


\maketitle


The formation of lanes is a ubiquitous phenomenon occurring in nature when two species of particles are driven against each
other. When the driving forces are strong enough, like-driven particles form ``stream lines'' and move collectively in
lanes. Typically, the lanes exhibit a considerably anisotropic structural order accompanied by an enhancement of their
mobility. The phenomenon is most commonly known from pedestrian dynamics in highly populated pedestrian zones
\cite{helbing_prl_2000}, but also occurs in very different model systems of driven particles, such as colloidal dispersions
\cite{dzubiella_pre_2002,leunissen_nature_2005,Reichhardt2006}, lattice gases \cite{Zia} and molecular ions
\cite{netz_epl_2003}. Lane formation is a genuine nonequilibrium transition \cite{Zia} which depends on the details of the
particle interactions and their dynamics \cite{Rex2008}.

Recently, the particle laning was also observed in complex plasmas \cite{Morfill06NJP}. In fact, complex plasmas
\cite{Fortov05,ShuklaBook} provide a very important intermediate dynamical regime that is between classic undamped fluids
and fully damped colloidal suspensions: In complex plasmas, the ``atomistic'' dynamics associated with the interparticle
interaction is virtually undamped whereas the large-scale hydrodynamics is determined by friction.

In this Letter we report on comprehensive experimental studies of lane formation in complex plasmas, that were carried out
under microgravity conditions with the radio frequency (rf) discharge chamber PK-3 Plus \cite{Thomas08}. The motivation for
this research is threefold: First, we demonstrate that complex plasmas are indeed an ideal model system to study
nonequilibrium phase transitions such as laning. Second, the experiments enable us to investigate the dynamical onset of
lane formation in detail. Third, we achieve a quantitative understanding of the structural correlation during the onset of
laning by comparison with particle-resolved Langevin simulations. Based on the anisotropic scaling index analysis of the
obtained data, we suggest a universal order parameter for non-equilibrium phase transitions in driven systems.


\begin{figure}
\includegraphics[width=7.cm,clip=]{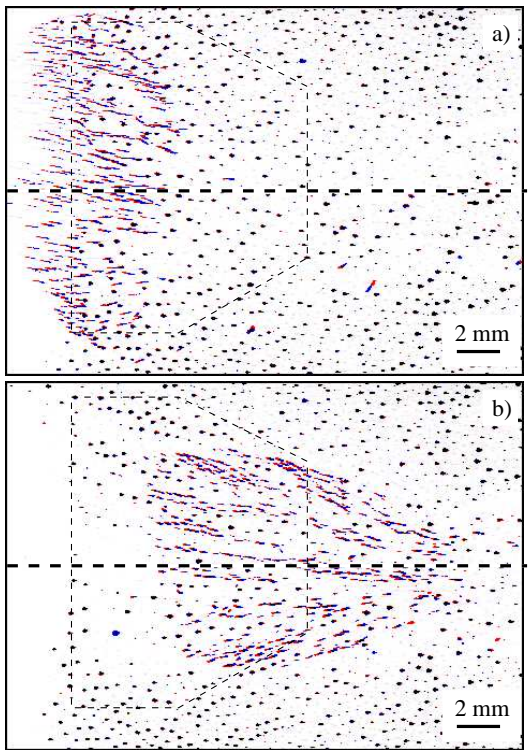}
\caption{\label{PKE_lane} Lane formation in complex plasmas. A short burst of small ($3.4~\mu$m) particles is injected into
a cloud of big ($9.2~\mu$m) background particles (close to the midplane of the chamber, indicated by horizontal dashed line).
Small particles are driven towards the center, stages of (a) initial lane formation and (b) merging of lanes into larger
streams are shown. Particles are illuminated by a thin laser sheet of $\simeq0.35$~mm; each figure is a superposition of
two consecutive color-coded images (1/50th s apart, red to blue), the time difference between them is $\simeq1.2$~s. At the
stage (b) big particles also form well defined lanes that can be identified as strings of red dots. The frame indicates
the region used for the analysis of big-particle dynamics. }
\end{figure}

{\it Experiments.} A series of dedicated experiments was carried out in the PK-3 Plus rf discharge chamber on the
International Space Station. Microgravity is a crucial factor which allows us to form stable spheroidal complex plasma
clouds. When small particles are injected into a cloud of big ones, the force field pulls small particles towards the
center, thus making such systems perfectly suited to study lane formation. Our experiments were performed for various
combinations of ``big'' and ``small'' monodisperse particles (2.55, 6.8, 9.2, and 14.9~$\mu$m diameter for ``big'', and
1.55, 2.55, and 3.4~$\mu$m for ``small''), with different neutral gases and pressures (argon between 10 and 60~Pa and neon
at 60~Pa, to control the friction rate), and for different rf discharge powers (to control the screening length).

Figure~\ref{PKE_lane} shows a characteristic example of lane formation observed in experiments with $3.4~\mu$m and
$9.2~\mu$m particles at pressure of 30~Pa. When a fraction of individual small particles enters the interface of the fairly
homogeneous cloud formed by big particles, the subsequent penetration is accompanied by a remarkable self-organization
sequence: (a) Big particles are pushed collectively by the inflowing cloud of small particles, the latter form strings
drifting on average along the force field. (b) As the particles approach the center of chamber, the field decreases and the
strings organize themselves into larger streams. At the later stage, when the field almost vanishes, the streams merge to
form a spheroidal droplet with well-defined surface, indicating the transition to the regime when surface tension plays the
primary role. In this Letter we are specifically interested in the first two stages \cite{footnote1}. It is noteworthy that
during the stage (b) big particles also form well defined strings. Small and big particles create an ``array'' of
interpenetrating strings, where small particles are drifting to the right and big ones are almost immobile. After the flux
of small particles is exhausted, the big-particle strings slowly dissolve \cite{footnote2}.



{\it Computer simulations.}
First, the distribution of the characteristic parameters of the discharge plasma, such as plasma density, electron
temperature, and electric fields, was deduced from 2D simulations of the PK-3 Plus discharge chamber. We used the plasma
fluid code (described in Ref.~\cite{Land06}), which provides a self-consistent coupling of dust species to the discharge
plasma, including the particle charging, plasma absorption, redistribution of volume charges/ambipolar fields, etc. These
simulations suggest that in the midplane of the chamber the cloud of big particles is ``self-confined'' at its edges due to
the self-consistent plasma field, as illustrated in Fig.~13 of Ref.~\cite{Land06}, whereas inside the cloud the electric and
ion drag forces practically compensate each other. On the other hand, since these two forces have different scaling on the
particle size, the net force on small particles is nonzero inside the cloud. For the experiment shown in
Fig.~\ref{PKE_lane}, the simulations yield a force of $f_{\rm s}=0.3\pm0.1$~pN pushing a small particle towards the center.

\begin{figure}
\includegraphics[width=7.cm,clip=]{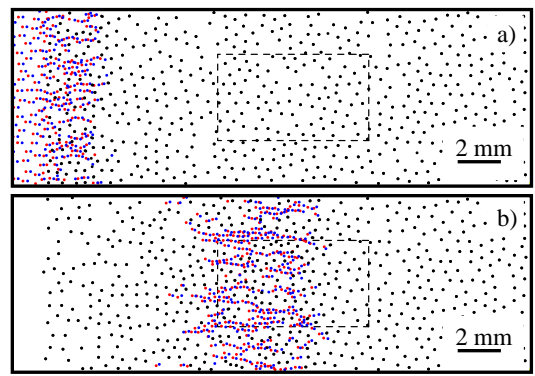}
\caption{\label{MD_lane} Lane formation in MD simulation corresponding to the experiment shown in Fig.~\ref{PKE_lane}.
Two snapshots illustrate (a) the initial injection stage and (b) the steady state.}
\end{figure}

Next, particle-resolved molecular dynamics (MD) simulations were performed on the Langevin level \cite{Ivlev05} for a binary
mixture of 5759 small and 12287 big particles. The simulation box with periodic boundary conditions has dimensions of
$4.4$~cm in $x$-direction and $0.8$~cm in $y$- and $z$-directions. The particles interact via a Yukawa pair potential with a
screening length $\lambda=100~\mu$m (based on results of our plasma discharge simulations) and effective charges $Z_{\rm
s}=3000~e$ (based on experiment \cite{Annibaldi07}) and $Z_{\rm b}=8117~e$, proportional to the respective particle
diameter, $\sigma_{\rm s}=3.4~\mu$m and $\sigma_{\rm b}=9.2~\mu$m. The mass density of the particles is $1.5~$g/cm$^3$ and
the corresponding friction rates are $\nu_{\rm s}=250$~s$^{-1}$ and $\nu_{\rm b}=(\sigma_{\rm s}/\sigma_{\rm b})\nu_{\rm
s}=92.4$~s$^{-1}$. The mean interparticle distances are deduced from the experiment, $\Delta_{\rm s}=464~\mu$m (before the
penetration) and $\Delta_{\rm b}=493~\mu$m, the temperature is $T=0.024$~eV.

%
%


Our plasma simulations show that big particles do not experience an external force in the bulk. Therefore, in our MD
simulation we confine them in $x$-direction by a parabolic external potential at two edges (with a width of 2.2~cm) and
adjust the confinement strength, so that the measured interparticle distance $\Delta_{\rm b}$ is reproduced. Similarly, a
portion of small particles, separated from the big particles, was prepared (with a width of 0.7~cm). Then the constant
driving force $f_{\rm s}$ was instantaneously applied, leading to penetration of small particles into the cloud of big ones.
Simulation snapshots are shown in Fig.~\ref{MD_lane}, revealing a qualitative agreement with the experiment.



{\it Anisotropic scaling index and order parameter.} In order to identify and quantify the string-like structures in our
experimental data and the simulations, a suitable order parameter has to be employed that is sensitive to the changing
particle structures. Conventional approaches, e.g., binary correlation or bond orientation functions, Legendre polynomials,
etc.\ turned out to be too insensitive. Much more satisfactory results were obtained by implementing an {\it anisotropic
scaling index} method -- a local nonlinear measure for structure characterization. This method has already been used to
characterize electrorheological complex plasmas \cite{Ivlev08}, large-scale distribution of galaxies \cite{Rath02}, or bone
structure \cite{Rath08}.


For a given set of particle positions, $\{{\bf r}_{i}\}$, $i=1,\ldots N$, we define a local density $\rho({\bf r}_{i},R) =
\sum_{j=1}^{N}s(d_{ij}/R)$, where $d_{ij}=|{\bf r}_{i}-{\bf r}_{j}|$ and $s$ is a certain shaping function characterized by
the spatial scale $R$. The scaling index $\alpha$ is the logarithmic derivative of the density with respect to the spatial
scale, $\alpha=\partial\log\rho({\bf r}_{i},R)/\partial\log R$. Hence, $\alpha({\bf r}_{i},R)$ characterizes the
dimensionality of the local structure around point ${\bf r}_{i}$, in the vicinity determined by the scale $R$. For example,
$\alpha({\bf r}_{i},R) \simeq 1$ means that the local structure is close to a straight line at the spatial scale $R$, for
$\alpha({\bf r}_{i},R) \simeq 2$ it is an element of a plane, and so on. Using the Gaussian shaping function
$s=e^{-(d_{ij}/R)^{2}}$ we derive for the scaling index,
\begin{equation}
\label{alpha}
\alpha({\bf r}_{i},R)=\frac{2\sum_{j=1}^{N}(d_{ij}/R)^{2}e^{-(d_{ij}/R)^{2}}}{\sum_{j=1}^{N}e^{-(d_{ij}/R)^{2}}}.
\end{equation}
Thus, the spatial scale $R$ is an important mesoscopic measure of the local environment. In the range of relevant scales
$\alpha$ is practically independent of $R$.


In order to characterize anisotropic structures, we use a ``stretch metric'' for the distance measure $d_{ij}$. On a 2D
plane, the metric is determined by the aspect ratio $\epsilon~(>1)$, which is the relative stretching of two principal axes,
and by the unit vector ${\bf u}= (\cos\theta,\sin\theta)$ in the direction of stretching. Then the resulting anisotropic
scaling index, $\alpha({\bf r}_{i},R,\theta)$, can be directly obtained from Eq. (\ref{alpha}).
We propose a ``uniaxial vector characterization'': Each point ${\bf r}_{i}$ is associated with the unit vector ${\bf u}_i$
which points to a ``preferred'' direction of the local anisotropy. This direction is determined by the angle $\theta_i$ at
which the ``anisotropic contrast'' $\alpha({\bf r}_{i},R,\theta_i+\pi/2)-\alpha({\bf r}_{i},R,\theta_i)$ is maximized. The
directions ${\bf u}_i$ and $-{\bf u}_i$ are equivalent, so that below they are defined for the range
$-\frac{\pi}2\leq\theta_i\leq \frac{\pi}2$.

Thus, each point can now be considered as a uniaxial ``molecule'' (simple rod) with the direction ${\bf u}_i$. Therefore,
the global laning on a 2D plane can be characterized with the second-rank tensor $Q_{\alpha\beta}=2N^{-1}\sum_{i=1}^{N}{\bf
u}_i\otimes{\bf u}_i-\delta_{\alpha\beta}$, analogous to that used to quantify order of the nematic phase. The direction of
the global laning, $\langle{\bf u}\rangle$, is then the eigenvector (``nematic director'') corresponding to the largest
eigenvalue of $Q_{\alpha\beta}$, which in turn is the laning order parameter, $S$. Obviously, $S=1$ for a perfect alignment
and $S=0$ for a disordered phase, when individual vectors ${\bf u}_i$ are uncorrelated. We finally define the global laning
angle $\Theta$ via the relation $\cos\Theta=\langle{\bf u}\rangle\cdot{\bf e}_x$.



{\it Comparison of experiments and simulations.} For the analysis of the MD simulations, we divided the volume of the
simulation box in $z$-direction (perpendicular to the driving force in $x$-direction) into several slabs of $0.35$~mm width,
which is about the thickness of the laser sheet used to record the experimental data. Hence, the average number of particles
and their average density, as well as the magnitude of their fluctuations in this ``reduced'' simulation data set were
similar to those in the experiment. The obtained results were analyzed and compared using the anisotropic scaling index
method. The relevant range of spatial scales $R$ was about $(2.5-4)\Delta_{\rm b}$ for big particles and $(3-5)\Delta_{\rm
b}$ for small particles, the anisotropic aspect ratio $\epsilon$ was 5 and 7, respectively.

Discrimination of big and small particles in the experimental data was performed in terms of their velocities: During the
stages of the lane evolution illustrated in Fig.~\ref{PKE_lane}, small particles drift relatively fast with respect to big
particles towards the center of the chamber. They gradually slow down, until at the later stage their velocities become too
low and then small and big particles are no longer distinguishable.

{\it (i) Small particles.} We first identified boundaries of the self-contained big- and small-particle clouds. Overlap of
the two clouds was a natural choice for the region to analyze the dynamics and structural evolution of small particles, as
they penetrate the cloud of big ones. There are two distinct phases characterizing formation and evolution of small-particle
lanes [approximately correspond to Figs~\ref{PKE_lane}(a) and \ref{PKE_lane}(b), respectively]: An ``injection stage'' (I),
starts at the moment when first small particles penetrate the cloud of big ones. During this phase, the number of small
particles for the analysis increases from zero to the average number per slab, and therefore significant fluctuations are
possible due to poor statistics. After about one second, there is a crossover to a ``steady-state stage'' (II), when the
average number of small particles remains constant and the driving force can be considered constant as well. In the
simulations the duration of steady-state phase is sufficiently long (two seconds), due to properly chosen length of the
simulation box. In the experiment, however, this phase is 2--3 times shorter, because the driving force and hence the
particle velocities decrease as they approach the center of chamber.

\begin{figure}
\includegraphics[width=6.4cm,clip=]{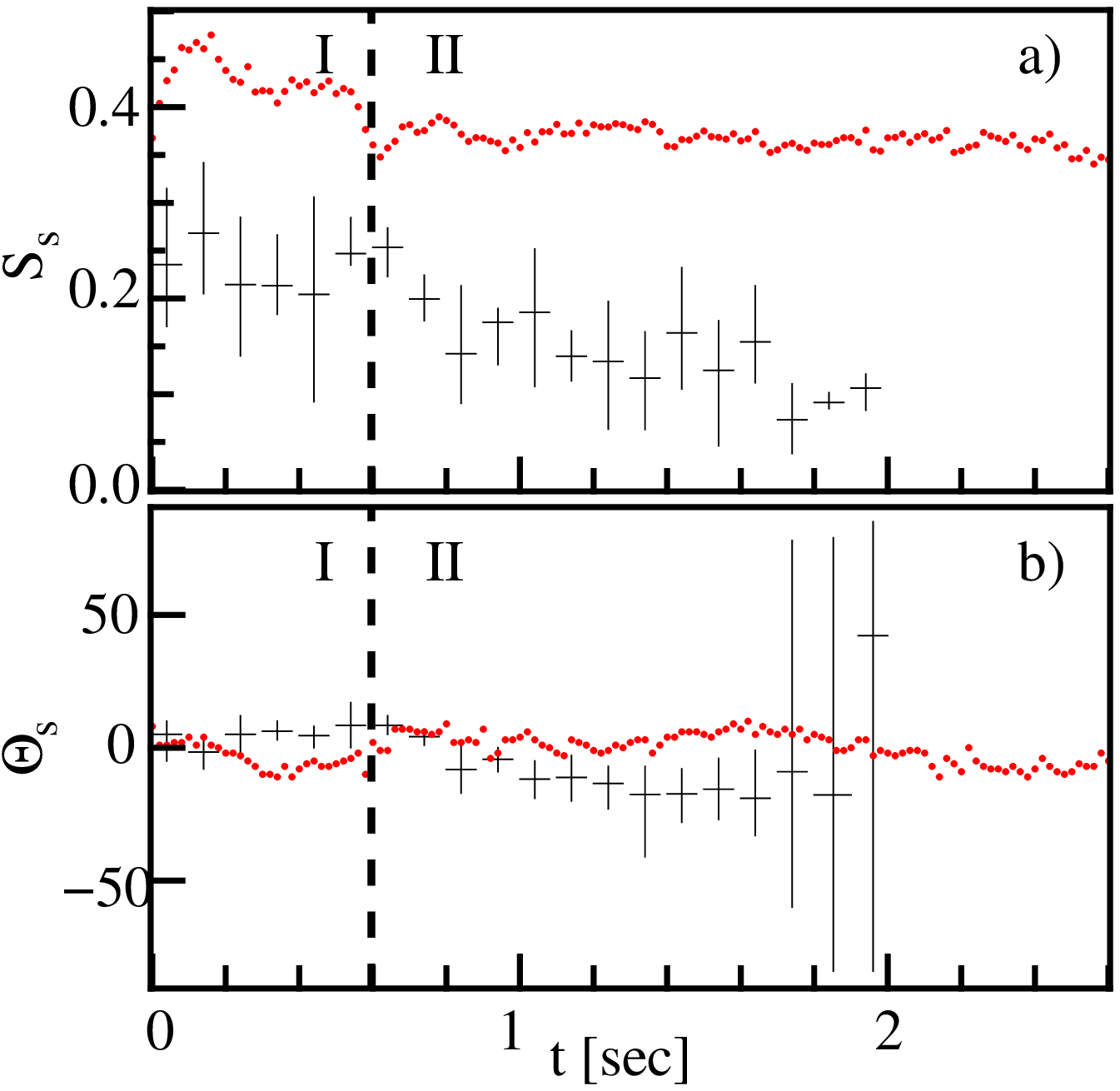}
\includegraphics[width=6.4cm,clip=]{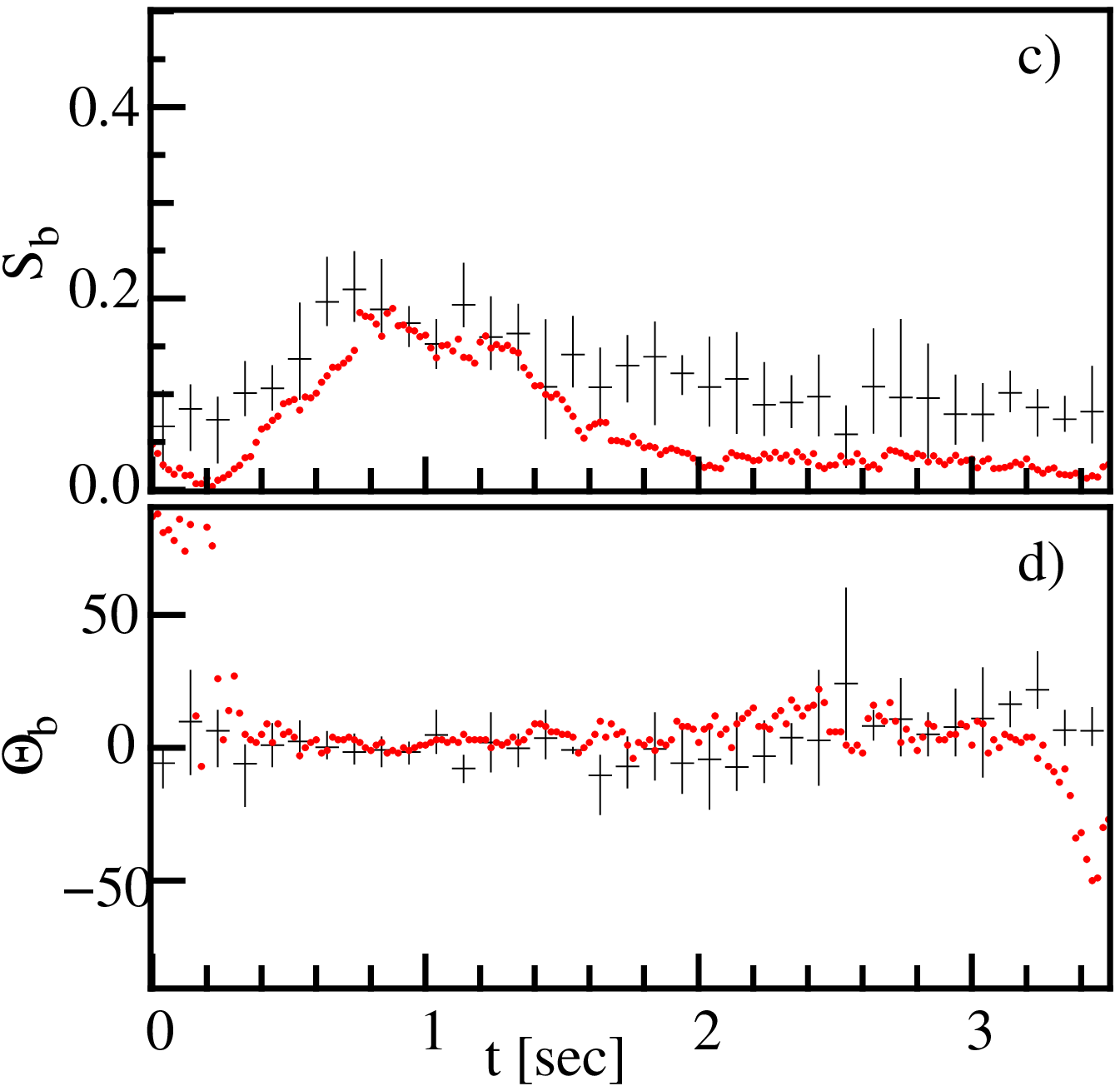}
\caption{\label{analysis} Dynamics of laning. Shown are the evolution of the ``nematic'' order parameter for small and
big particles, $S_{\rm s}$ (a) and $S_{\rm b}$ (c), respectively, as well as the corresponding global laning angle,
$\Theta_{\rm s}$ (b) and $\Theta_{\rm b}$ (d), as obtained from the anisotropic scaling index analysis of the experiment
(crosses) and MD simulation (dots). For small particles, injection stage I and steady-state stage II are indicated.}
\end{figure}

The order parameter $S_{\rm s}(t)$ calculated for small particles is plotted in Fig.~\ref{analysis}(a). Although in the
beginning of stage I it exhibits significant fluctuations due to poorer statistics, one can see that the formation of
small-particle lanes is practically ``instantaneous'' at timescales corresponding to the experimental frame rate (50
frames/s). The magnitude of the order parameter in the experiment is almost twice as low as that in the MD simulation. We
believe that this discrepancy is due to the fact that the discrimination procedure allows us to identify only 70--80~\% of
small particles in the experimental data, which results in the artificial ``thinning'' of the small-particle lanes. The
random elimination of 30~\% small particles in the MD data decreases $S_{\rm s}$ down to the experimental level, clearly
supporting this hypothesis. Figure~\ref{analysis}(b) shows the evolution of the global laning angle $\Theta_{\rm s}(t)$ for
small particles, which exhibits narrow dispersion and demonstrates that the ``nematic director'' practically coincides with
the driving vector. Note the increasing broadening and deflection of $\Theta_{\rm s}(t)$ seen in experiment at the
``steady-state'' stage II [coinciding with slight decrease in $S_{\rm s}(t)$], which is related to convergence of
small-particle lanes towards the chamber center as the driving force gradually decreases.

{\it (ii) Big particles.} In order to diminish possible influence of the boundary effects, we defined fixed regions in the
bulk of the big-particle cloud, as indicated in Figs~\ref{PKE_lane} and \ref{MD_lane}. In terms of the size and shape, these
regions approximately correspond to the ``overlap regions'' used for the analysis of small-particle dynamics.

The evolution of the order parameter and of the global laning angle for big particles is shown in Figs~\ref{analysis}(c) and
\ref{analysis}(d), respectively. Initially, there is no anisotropy in the simulation, whereas in the experiment $S_{\rm
b}\simeq0.05$ and $\Theta_{\rm b}\to0$, due to a weak inhomogeneity in the big-particle density. Once a cloud of small
particles reaches the fixed region used for the analysis, $S_{\rm b}$ starts growing and the angular distribution narrows
around $\Theta_{\rm b}=0$, due to the increasing number of small particles causing the formation of big-particle strings.
Then $S_{\rm b}$ reaches a maximum and starts falling off, reflecting the onset of string relaxation when small particles
leave the region. The initial relaxation occurs at a characteristic timescale of $\sim1$~s which is an order of magnitude
shorter than the self-diffusion timescale for the big particles ($\sim m_{\rm b}\nu_{\rm b}\Delta_{\rm b}^2/T$). Note that
after this rapid relaxation $S_{\rm b}(t)$ tends to some intermediate plateau, and laning angle $\Theta_{\rm b}(t)$ keeps
the anisotropy, indicating that the structural relaxation is apparently not complete. This suggests that the ultimate
equilibration might involve some metastable states.


Thus, binary complex plasmas provide us with crucial new insights into very important dynamical regime of laning, mediating
classic undamped fluids and fully damped colloidal suspensions. By combining the experimental studies and the
particle-resolved Langevin simulations, we investigated the dynamical onset of lane formation in driven complex plasmas.
Furthermore, we proposed a universal order parameter
for the characterization of non-equilibrium phase transitions in driven systems. The approach is based on the anisotropic
scaling index analysis that is exceptionally sensitive to symmetry changes occurring in particle ensembles. The use of such
order parameter could be invaluable in studying the onset of non-equilibrium phase transitions. In particular, this might
help us to shed the light into the principal issue about the order of phase transition, characterize possible universality,
identify dynamical regimes of structural relaxation, etc. As the immediate next steps, one can think of employing the
proposed approach to investigate laning in periodically driven \cite{Corte08}
and crystalline \cite{Roichman07} systems.

We thank J. Dzubiella, R. Monetti, and M. Rex for helpful discussions. This work was supported by DLR/BMWi (grant no.
50WP0203), SFB TR6, and RFBR (grant no. 08-02-00444).


\begin{thebibliography}{99}
\bibitem{footnote1} Droplet formation is governed by different physical mechanisms and will therefore be considered
    elsewhere.
\end{thebibliography}
\end{document}